\documentclass[aip,pra,reprint,amsfonts,amsmath,amssymb,superscriptaddress,10pt,floatfix]{revtex4-1}

\usepackage[english]{babel}
\usepackage[utf8]{inputenc}

\usepackage{amsmath}
\usepackage{amsfonts}
\usepackage{amssymb}
\usepackage{amsthm}
\usepackage{mathtools}
\usepackage{graphicx}
\usepackage{xfrac}

\usepackage{color}
\usepackage{array}
\usepackage{longtable}
\usepackage{calc}
\usepackage{multirow}
\usepackage{hhline}
\usepackage{ifthen}

\graphicspath{{figures/}}

\begin{document}

\title{Folding mechanism of a polymer chain with short-range attractions}

\author{Christian Leitold}
\author{Christoph Dellago}

\affiliation{Faculty of Physics, University of Vienna, Sensengasse 8, 1090 Vienna, Austria}

\date{September 15, 2014}

\begin{abstract}
We investigate the crystallization of a single, flexible homopolymer chain using transition path sampling (TPS). The chain consists of $N$ identical spherical mo\-no\-mers evolved according to Langevin dynamics. While neighboring mo\-no\-mers are coupled via harmonic springs, the non-neighboring monomers interact via a hard core and a short-ranged attractive potential. For a sufficiently small interaction range $\lambda$, the system undergoes a first-order freezing transition from an expanded, disordered phase to a compact crystalline state. Using a new shooting move tailored to polymers combined with a committor analysis, we study the transition state ensemble of an $N=128$ chain and search for possible reaction coordinates based on likelihood maximization. We find that typical transition states consist of a crystalline nucleus with one or more chain fragments attached to it. Furthermore, we show that the number of particles in the crystalline core is not well suited as a reaction coordinate. We then present an improved reaction coordinate, which includes information from the potential energy and the overall crystallinity of the polymer.
\end{abstract}

\maketitle

\section{Introduction}

In reaction to changes in their environment, polymers often go through large-scale conformational changes akin to phase transitions, which are of significance to many biological processes. A simple homopolymer chain, for instance, collapses from an extended coil to a compact globule in response to changes in the solvent~\citep{degennes_scaling}. This transition is continuous and can be viewed as the single chain analogy of a second-order phase transition. In other cases, the conformational change of the polymer occurs discontinuously in a first-order like transition, and distinct conformations can coexist with each other. Such cooperative behavior, familiar from two-state protein folding, has been also observed in a single polymer chain with square-well interaction~\cite{tpb_2}.

The complex phase behavior of the square-well chain has been investigated previously in a number of different studies. Taylor and Lipson~\cite{taylor_ex1, taylor_ex2} carried out analytical calculations on very small chains, and showed that the system's radius of gyration is a sigmoidal function of temperature. Later, Zhou~et\,al.~\cite{zhou} performed simulations of longer chains showing that the increase in the polymer's dimensions with temperature is more pronounced for longer chains. More recently, Taylor, Paul and Binder~\cite{tpb_2, tpb_1, tpb_4} mapped out the entire phase diagram as a function of temperature and width of the attractive well~\cite{tpb_1}. Depending on the well width, one observes one of two phase transitions: a second-order coil-to-globule for wide wells, and a first-order coil-to-crystal transition. The authors also calculated the free energy as a function of the potential energy at coexistence. In the latter case a free energy barrier separating the two phases imposes a pronounced two-state behavior on the system~\cite{tpb_4}.

In this work, we focus on clarifying the mechanism of the coil-to-crystal transition of the polymer. In order to perform molecular dynamics simulations and hence obtain a realistic picture of the transition event, we have developed a continuous approximation to the pure square-well chain. Using this model we performed transition path sampling (TPS) simulations~\cite{tps0, tps1, tps2} to obtain reactive pathways taking the chain from the expanded to the folded configuration. To enhance the efficiency of our TPS simulation, we devised a new shooting move where the topology of the polymer is altered prior to the actual shooting. We then calculated committor values for the configurations along the transition pathways and used this information as a basis for the search for a possible reaction coordinate based on the likelihood maximization method of Peters and Trout~\cite{peters_rc}. An analysis of the transition state ensemble (TSE) of the system indicates that the number of crystalline particles in a given configuration is a rather poor measure of the progress of the transition. Similarly, the chain's radius of gyration does not convey any information about the folding transition. However, a combination of the potential energy with the global order parameter $Q_6$ improves the quality of the reaction coordinate.

The remainder of this paper is organized as follows. In Sec.\,\ref{sec:model} we define the model and summarize its properties. Section\,\ref{sec:methods} covers the numerical methods employed and the details of our simulations. Results are presented in Sec.\,\ref{sec:results}, and a discussion is provided in Sec.\,\ref{sec:discussion}. Details of the new TPS shooting move are given in the Appendix.

\section{Polymer model}
\label{sec:model}

\subsection{Square-well chain}

\begin{figure}
\includegraphics[width=0.95\columnwidth]{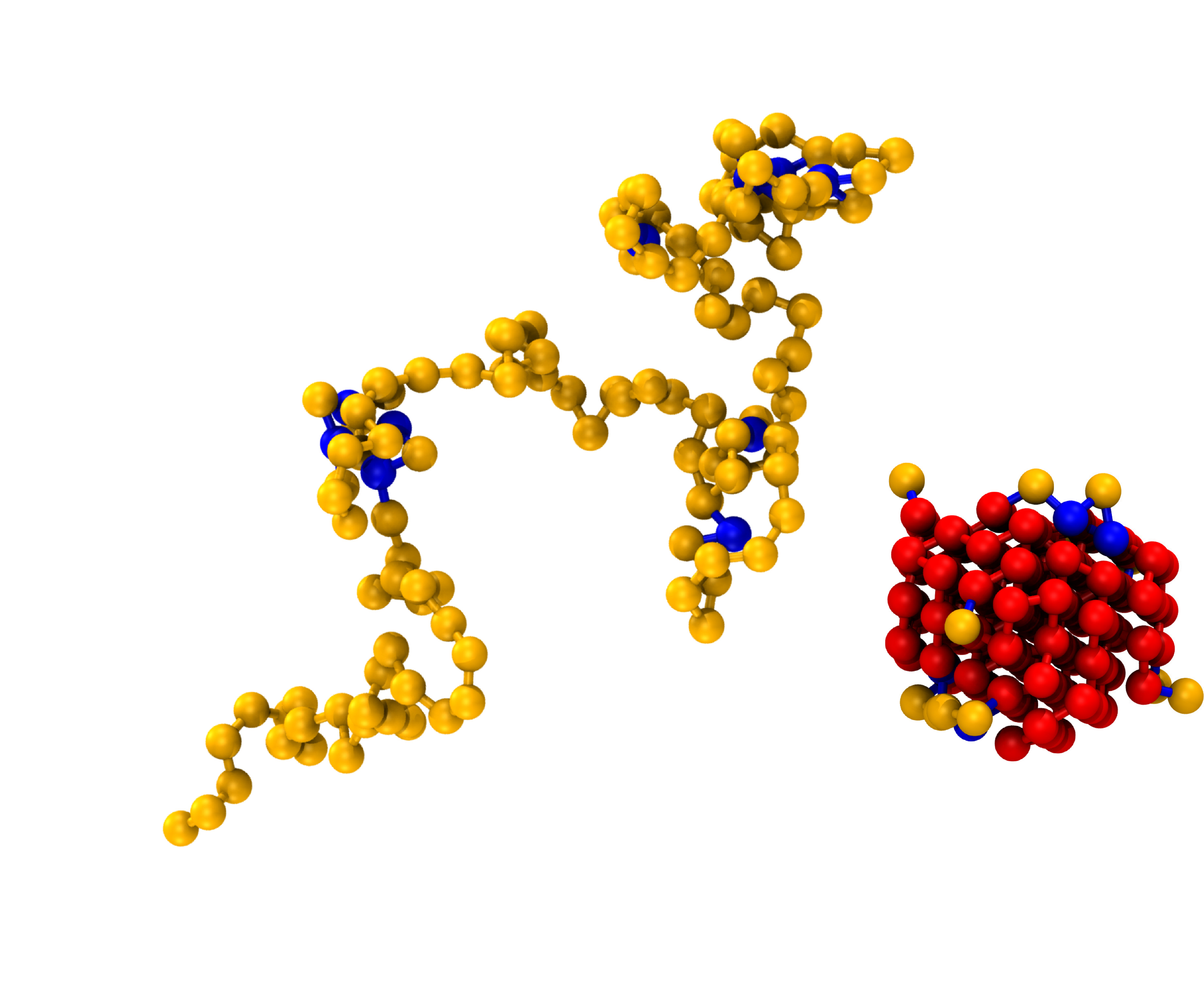}
\caption{Coil (left) and crystalline (right) state of the polymer chain for particle number $N=128$, interaction range $\lambda = 1.05$ and temperature $k_{\rm{B}} T / \varepsilon =0.438$. Crystalline and coil-like particles are colored in red and yellow, respectively, while intermediate particles are colored in blue. The criterion for crystallinity used here is defined in Sec.\,\ref{sec:methods}.}
\label{fig:states}
\end{figure}

The square-well polymer model considered previously in the work of Taylor, Paul and Binder~\cite{tpb_1, tpb_2, tpb_4} is a fully flexible chain of $N$ identical monomers. While the distance $\sigma$ between neighboring monomers is fixed,  non-neighboring monomers interact via the square-well pair potential~\cite{yethiraj_hall}
\begin{equation}
u(R_{ij}) = \begin{cases}
\infty  & 0 \leq R_{ij} < \sigma, \\
- \varepsilon & \sigma \leq R_{ij} \leq \lambda \sigma, \\
0 & R_{ij} > \lambda \sigma.
\end{cases}
\label{squarewell}
\end{equation}
Here, $R_{ij} = | \vec{R}_i -\vec{R}_j |$ is the distance between the $i$-th and $j$-th monomer and $\lambda > 1$ parametrizes the width of the potential well. Due to the form of the potential, the system has a discrete energy spectrum $E_n = - n \varepsilon$, where $n$ is the number of square-well overlaps in the chain.

As recently shown by Taylor, Paul and Binder~\cite{tpb_2,tpb_1}, depending on the value of the potential width $\lambda$ and temperature $T$, there exist three different phases. At high temperatures, the polymer is in the expanded coil phase for all values of $\lambda$. What happens if the system is cooled, however, depends on the width $\lambda$ of the attractive well. For wide wells, upon cooling the chain first collapses into a compact but disordered globule. This transition is of second order. Further lowering the temperature then leads to a first-order phase transition to a crystalline state. For sufficiently narrow wells ($\lambda \leq 1.05$), the system directly freezes from the expanded coil to the crystalline state in a first-order transition. Snapshots of the coil state and the crystal state are depicted in Fig.\,\ref{fig:states}.

\subsection{Chain with smoothed square well}

\begin{figure}
\includegraphics[width=0.95\columnwidth]{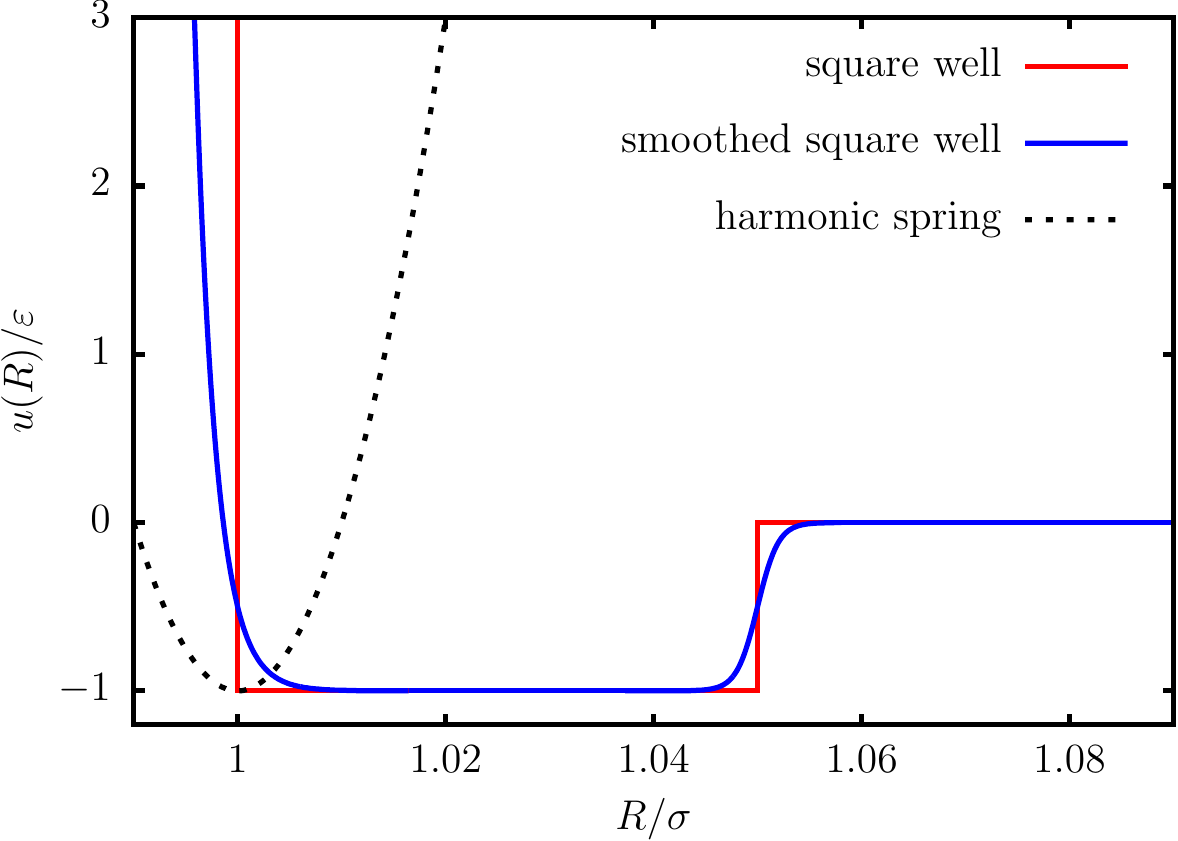}
\caption{Original square-well and new smoothed potential for $\lambda = 1.05$. The harmonic spring potential (acting between neighboring beads) is also shown for comparison.}
\label{fig:potentials}
\end{figure}

Since we are interested in the mechanism of the coil-to-crystal transition, the procedure used to evolve the system along transition pathways must resemble the natural dynamics of the system. If Monte Carlo dynamics is considered, this implies that only local moves can be used. Molecular dynamics provides a more physical (and computationally more efficient) way to model the time evolution of the system. To facilitate such simulations and avoid the cumbersome handling of impulsive forces caused by the discontinuities  in the potential, we have developed a smooth (differentiable) version of the square-well potential. The smooth potential is defined as
\begin{equation}
u(R) = \frac{\varepsilon}{2} \left\{ \exp \left[ \frac{-(R-\sigma)}{a} \right] + \tanh \left[ \frac{R-\lambda \sigma}{a} \right] - 1 \right\},
\end{equation}
where we have chosen a value of $a = 0.002 \, \sigma$ for the parameter which determines the steepness of the exponential repulsion and the width of the step at $R = \lambda \sigma$. Neighboring monomers are coupled via harmonic springs $U(R) = \frac{k}{2} (R-\sigma)^2$ with a value of $k= 20000 \, \sigma^2 / \varepsilon$ for the spring constant. A comparison of the original square-well potential and its smooth variant is shown in Fig.\,\ref{fig:potentials}. For our simulations, we have chosen the $N=128$ chain with an interaction length of $\lambda = 1.05$.

\section{Methods}
\label{sec:methods}

In this Section we provide a brief outline of the computational methods used to obtain the results discussed in Sec.\,\ref{sec:results}.

\subsection{Wang--Landau simulation}

In order to verify the equivalence of our smooth chain model with the square-well chain studied previously~\cite{tpb_1,tpb_2,tpb_4}, we have performed Wang--Landau simulations~\cite{wang-landau}. In the Wang--Landau algorithm, one iteratively obtains the density of states $g(E)$, from which all other thermodynamic properties follow. This is done by performing a Monte Carlo simulation with the inverse of the current estimate of the density of states as the weight of a configuration, instead of the usual Boltzmann weight. A single Monte Carlo move is one of five distinct possibilities, selected at random~\citep{tpb_1}:
\begin{enumerate}
\item pivot move: a monomer $i$ within the chain is selected randomly and the whole chain segment $[i+1,N]$ undergoes a random Euler rotation about the $i$-th monomer;
\item end move: the 1st or $N$-th monomer is rotated by a small angle about a random axis;
\item crankshaft move: a randomly chosen monomer within the chain is rotated by a small angle about the axis defined by the line connecting the two neighboring particles;
\item bond-bridging move: described in Sec.\,\ref{sec:core-modification};
\item standard displacement move: a randomly chosen monomer is moved by a small random displacement.
\end{enumerate}
One pass of the Wang--Landau simulation consists of $2N$ single moves and contains, on average, $2$~pivot moves, $2$~end moves, $N/2 - 4$ crankshaft moves, $N/2$~bond-bridging moves and $N$ displacement moves. For the energy histogram $H(E_n)$, we have used bins of unit width in an energy range of $[-430 \, \varepsilon, 400 \, \varepsilon]$, which in order to speed up the computations is split into two overlapping energy ranges. The results are then averaged and joined at $E_{\rm{join}} = -190 \, \varepsilon$. The quality of the joint is checked by computing the numerical derivative of the density of states, which is found to agree well in the vicinity of $E_{\rm{join}}$. The histogram  $H(E_n)$ is checked every $10^4$ passes for flatness, and is considered flat if no bin deviates by more than 20\,\% from the average. Furthermore, as an alternative flatness criterion, we check for uniform growth of $H(E_n)$ (i.\,e., each bin in the histogram has increased within a range of 20\,\% of the average growth since the last check) every $5 \times 10^7$ passes. To keep the numbers within the range that can be handled by standard floating-point arithmetic, $\ln g(E)$ instead of $g(E)$ is stored. Our initial value of the density of states is $g(E_n) = 1$ for all $E_n$, with an initial modification factor of $f_0 = e^1$. Subsequent modification factors are chosen as $f_{m+1} = \sqrt{f_m}$.

Once the density of states is known, one can calculate the canonical partition function
\begin{equation}
Z(T) = \sum_n g(E_n) e^{-\beta E_n}
\end{equation}
and the probability density of the energy,
\begin{equation}
P(E_n, T) = \frac{1}{Z(T)} g(E_n) e^{-\beta E_n}.
\end{equation}
Here, $\beta = 1 / k_{\rm{B}} T$ is the inverse temperature. With that, any quantity which is a function of the energy can be calculated, in particular the specific heat
\begin{equation}
C(T) = \frac{1}{k_{\rm{B}} T^2} ( \langle E^2 \rangle - \langle E \rangle^2), 
\end{equation}
where
\begin{equation}
\langle E \rangle = \sum_n E_n P(E_n, T).
\end{equation}
For a finite system, maxima in $C(T)$ are related to phase transitions or other structural rearrangements~\cite{zhou_hall_karplus}.

\subsection{Transition path sampling with core modification}
\label{sec:core-modification}

In order to gain insight into the transition mechanism, we have performed transition path sampling simulations for a temperature at which the coil and crystal coexist. At this particular temperature, the equilibrium probability distribution of the energy is bimodal with two peaks of equal weight for the coil and crystallite, respectively. In between, the probability is extremely low, which corresponds to a high free energy barrier separating the two types of configurations. Hence, in an equilibrium molecular dynamics or Monte Carlo simulation, one would practically never see a direct transition from the coil to the crystallite or vice versa. We overcome this time scale problem by using transition path sampling. In contrast to Wang--Landau sampling, TPS is able to provide dynamical pathways taking the system from one state to another.

Our TPS simulations use aimless shooting~\citep{peters_rc} and a flexible path length. We employ Langevin dynamics with a time step $\Delta t = 0.0002 \, \sqrt{m \sigma^2 / \varepsilon}$ and a damping constant $\gamma = 0.5 \, m^{3/2} \sigma^2 \varepsilon^{-1/2}$, where the integration of the equation of motion is performed using the Langevin thermostat by Schneider and Stoll~\cite{schneider} implemented in a modified version of LAMMPS~\cite{lammps}. The distinction between the states $A$ (expanded coil) and $B$ (frozen crystallite) is made based on the potential energy of the system. A configuration is considered to be in the coil state if $U/N \geq U_{\rm{min}}/N = -0.7 \, \varepsilon$ and it is considered to be in the crystalline state if $U/N \leq U_{\rm{max}}/N = -2.6 \, \varepsilon$. We check if one of these states is reached every $n_{\rm{check}} = 2000$ time steps, which is also the time interval for saving the current snapshot along the path. The shooting point is selected with equal probability from $x_0^{(o)}$, $x_{0+\Delta T}^{(o)}$ and $x_{0-\Delta T}^{(o)}$, where $x_0^{(o)}$ denotes the shooting point of the previous trajectory and $\Delta T = 50 \times n_{\rm{check}} \, \Delta t$. No restrictions are placed on the length of the pathways. A path is accepted if it is reactive, i.\,e., $A \rightarrow B$ or $B \rightarrow A$, and rejected if it is not.

One major problem when performing a standard transition path sampling simulation of the polymer chain is that the decorrelation between subsequent pathways is very slow. In particular, subsequent pathways generated with the standard shooting move have almost identical crystalline cores in the transition state region. A solution of this problem is to modify the shooting point prior to each shooting move more than by just modifying the particle velocities. In particular, at each shooting point we perform a number of bond-bridging Monte Carlo moves~\citep{tpb_1,escobedo_jacobian} in addition to a random assignment of the particle velocities. This bond-bridging moves modify also the crystalline core, leading to a faster decorrelation of the pathways.

In order to maintain the equilibrium distribution of states, the bond-bridging move has to be performed at the same temperature as the actual transition path sampling simulation. In this move, one first chooses the first ($i=1$) or last ($i=N$) monomer of the polymer with equal probability. Then, all interior monomers ($i > 3$ or $i < N - 2$) within a distance of $2 \, \sigma$ of the chosen end are identified. One of these monomers is chosen at random and then connected to monomer $1$ (or $N$) via removal of monomer $i-1$ (or $i+1$) and re-insertion between monomer $i$ and the selected end at a random azimuthal angle. The re-insertion is furthermore done such that none of the bonded distances changes. The move is then accepted with probability
\begin{equation}
P_{\rm{acc}}(a \longrightarrow b) = \min \left( 1, e^{-\beta \Delta U} \frac{b_a}{b_b} \frac{R_b}{R_a} \right),
\end{equation}
where the potential energy change $\Delta U = U_b - U_a$ stems only from the change in the potential for the non-neighboring sites, $b_a$ and $b_b$ are the numbers of possible bridging partners in the initial and final state, and $R_a = R_{1i}$ (or $R_{iN}$) is the distance between monomer $i$ and the selected end, with $R_b$ defined accordingly for the final state. A detailed description of the algorithm adapted to smooth bonding potentials is provided in Appendix~\ref{ap:bond_bridging}. It is worth noting that due to the bond-bridging moves carried out at the shooting point, the newly generated pathway does not intersect with the previous path at the shooting point. Still, each individual path is continuous in both space and momentum.

\subsection{Particle classification}

In order to distinguish between coil-like and solid particles, we use the connection coefficients $d_{ij}$ based on Steinhardt bond order parameters~\cite{steinhardt, tenwolde_bondorder}. For two particles $i$ and $j$ close to each other (distance smaller than $1.05 \, \sigma$), the connection coefficient $d_{ij}$ is defined as the scalar product of the complex $q_6$ vectors of particles $i$ and $j$ expressed in terms of spherical harmonics~\cite{steinhardt}. The scalar product measures the correlation between the local environments of particles $i$ and $j$. We then define two adjacent particles $i$ and $j$ as \textit{connected} if $d_{ij} \geq 0.5$.  Furthermore, let us denote the number of (bonded and non-bonded) neighbors and the number of connected neighbors of particle $i$ with $N_n(i)$ and $N_c(i)$, respectively. Particle $i$ is defined as \textit{crystalline} if $N_n \geq 5$ and $N_c \geq N_n - 1$. This combination of two conditions ensures that surface particles with a reduced number of adjacent particles can be detected as crystalline. At the same time, particles deep within the core of the polymer are not incorrectly detected as crystalline if they belong to the compact, but unordered phase. In addition, a particle is defined as \textit{coil-like} if $N_n \leq 4$ regardless of the value of $N_c$. We classify particles as \textit{intermediate} if they are neither crystalline nor coil-like.

\subsection{Overlap between configurations}

In order to investigate the effectiveness of the core-modification shooting move in generating new trajectories, we introduce a measure for the overlap between system configurations. The key idea is that two configurations are classified as similar if they share contacts between the same particles. Here, we use the matrix of pair energies as a measure of contact. Due to the shape of the pair potential, typically the pair energies in units of $\varepsilon$ are very close to $-1$ (in contact) or $0$ (not in contact). For two configurations, represented by contact matrices $X$ and $Y$, we define the overlap as
\begin{equation}
c_{XY} = \frac{ \sum_{i,j} X_{ij} Y_{ij}} { \sqrt{ \sum_{i,j} X_{ij}^2}  \sqrt{ \sum_{i,j} Y_{ij}^2} },
\label{eq:overlap}
\end{equation}
where $i$ and $j$ run over all monomers in the system. Note that by taking the element-wise product of the matrices, only connections which are present in both configurations give a contribution to the overlap measure. Also, the normalization corrects for the total number of contacts present and ensures that the overlap between two identical configurations is $1$. Taking this idea one step further, we define a correlation function, which measures the average overlap over a time series of configurations, for example the shooting points of a long TPS simulation,
\begin{equation}
c(n) = \frac{\langle \delta X(0) \cdot \delta X(n) \rangle}{\langle \delta X^2 \rangle}.
\end{equation}
Here, the matrix $\delta X(n) = X(n) - \langle X \rangle$ is the deviation of the contact matrix at step $n$ from its average value and the product is meant in the sense of Eq.\,\eqref{eq:overlap}. By definition, $c(0) = 1$.

\subsection{Committor analysis}

For a system with two (meta-) stable states $A$ and $B$, the committor $p_B(\mathbf{r})$ of configuration $\mathbf{r}$ is the fraction of dynamical pathways started from $\mathbf{r}$ that first reaches state $B$~\cite{tps0}. In practice, one launches a number of trajectories starting with random momenta from $\mathbf{r}$ and counts the fraction of trajectories ending in $B$. Our committor calculations were performed according to the algorithm described in Ref.\,\onlinecite{tps_fluid}, using $N_{\rm{min}} = 100$ and $N_{\rm{max}} = 500$.

\subsection{Reaction coordinate analysis}

To find a reaction coordinate capable of quantifying the progress of the transition we follow the likelihood maximization approach proposed by Peters and Trout~\cite{peters_rc}. In this method, a proposed reaction coordinate $r$ is modeled as a linear combination of $m$ physical parameters $q_k$:
\begin{equation}
r(\mathbf{q}) = \sum_{k=1}^{m} \alpha_k q_k + \alpha_0,
\label{eq:rofq}
\end{equation}
where $\mathbf{q} = \{ q_k \}$.
The committor $p_B$ is assumed to be a sigmoidal function of this model reaction coordinate,
\begin{equation}
p_B(r) = \frac{1}{2} [1+\tanh(r)].
\label{eq:pbofr}
\end{equation}
The coefficients $\alpha_k$ are then chosen such that the likelihood
\begin{equation}
L(\alpha) = \prod_{k}^{B} p_B(r^{(k)}) \prod_{k'}^{A} [1-p_B(r^{(k')})]
\label{eq:Lalpha}
\end{equation}
is maximized. In the above equation, the first product runs over all single shooting events ending in state $B$, while the second product runs over all shooting events ending in state $A$. The likelihood $L(\alpha)$ quantifies the compatibility of the proposed model with the measured committor values. We use the Bayesian information criterion~\cite{bic}
\begin{equation}
\mathrm{BIC} = {-2 \ln{L(\alpha)} + (m+1) \ln(n)}
\end{equation}
to compare the optimization results for different numbers of optimization parameters, where smaller BIC values are better. Here, $n$ is the total number of observations, i.\,e., the total number of shooting events entering in Eq.\,(\ref{eq:Lalpha}), and $m$ is the number of free parameters entering the model. The BIC penalizes models with too many free parameters, hence it is used to check whether it is sensible to add additional physical parameters to improve the model reaction coordinate.

To carry out the reaction coordinate analysis, we first calculate the committor for a number of states randomly selected from transition pathways. We then calculate a set of collective variables for each state. In order to obtain the optimal combination of collective variables, we first determine the single variable that maximizes the likelihood defined above. Then, we maximize the likelihood for each 2-variable combination of the selected variable and all the remaining other variables. The procedure is repeated successively adding variables. Combinations of up to 3~variables are considered.

For a perfect reaction coordinate $r(\mathbf{q})$ all configurations with the same value of $r$ should have the same committor, $p_B(\mathbf{q}) = p_B[r(\mathbf{q})]$~\cite{hummer_committor, vandeneijnden_committor}. In particular, there should exists a value $r^*$ of the reaction coordinate such that all configurations with $r^*$ have a committor of $p_B = 1/2$, i.\,e., they have equal likelihood to relax into state $A$ or $B$. These configurations are transition states and together they form the transition state ensemble (TSE). Accordingly, the distribution of committor values for states with $r(\mathbf{q}) = r^*$ should be strongly peaked around a value of $0.5$. For a perfect reaction coordinate, the only deviation from $p_B = 0.5$ is due to the statistical error that arises in computing the committor from a finite number of trajectories. Therefore, the width of the peak around $p_B = 0.5$ decreases with increasing number of trajectories used to calculate the committor for each configuration.

\section{Results}
\label{sec:results}

\subsection{Density of states and heat capacity}

\begin{figure}
\includegraphics[width=\columnwidth]{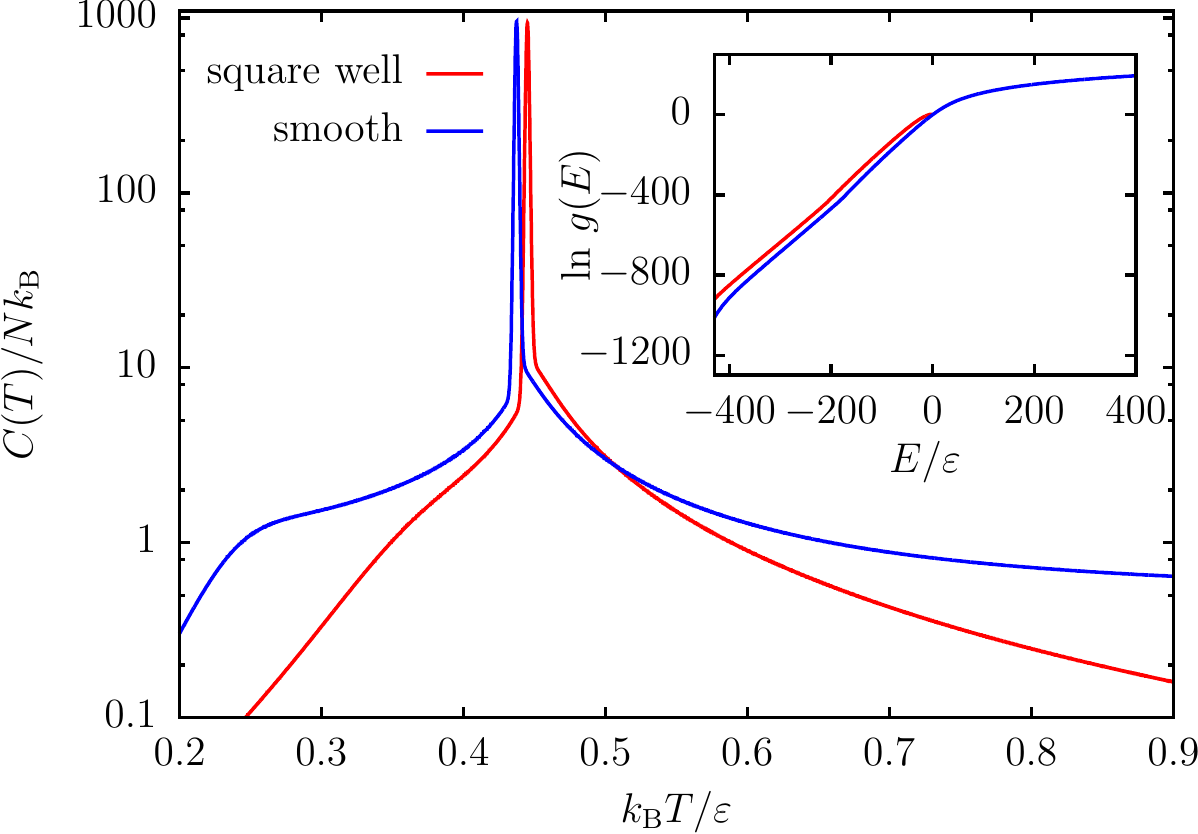}
\caption{Specific heat per monomer $C(T)/N k_{\rm{B}} T$ for the square-well model and its smooth modification. For the smooth chain, the freezing peak is located at $k_{\rm{B}} T / \varepsilon = 0.438 \pm 0.001$, while the value for the square-well chain is $k_{\rm{B}} T / \varepsilon = 0.446 \pm 0.001$, in accordance with the result from Taylor et\,al~\cite{tpb_1}. Inset: density of states $\ln g(E)$. Note that for the pure square-well chain, no positive energies are possible due to the absence of (finite) positive potentials. Both curves are normalized such that $\ln g(0) = 0$.}
\label{fig:UT}
\end{figure}

\begin{figure}
\includegraphics[width=\columnwidth]{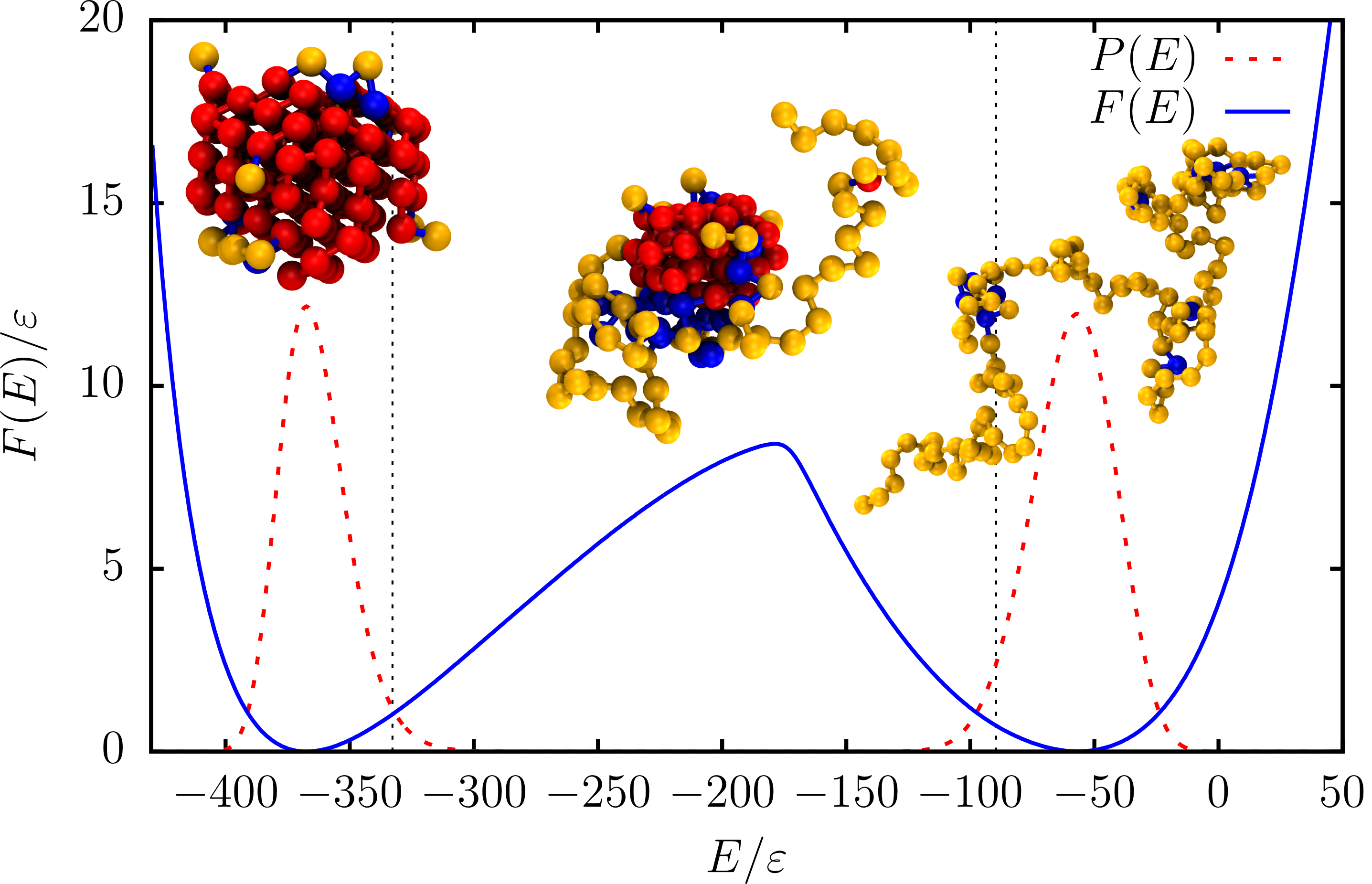}
\caption{Free energy (solid line) and probability distribution (dashed line) as a function the potential energy at the freezing temperature $k_{\rm{B}} T / \varepsilon = 0.438$. The two peaks correspond to the crystalline and coil state, respectively. The stable state boundaries used in our TPS simulations are indicated by dashed vertical lines. The configurations shown are snapshots of the coil (right) and crystalline (left) states and the top of the free energy barrier (center).}
\label{fig:FEPE}
\end{figure}

In order to relate the freezing transition of the smooth polymer chain to that of the pure square-well chain, we first need to verify that its phase behavior is similar to that of the original model. Hence, we have calculated the specific heat $C(T)$ for both the original square-well chain as well as for its smooth variant using the density of states (Fig.\,\ref{fig:UT}) obtained by Wang--Landau sampling. Two important observations can be made. First, the overall structure of the two curves is almost identical, especially in the important region corresponding to the freezing transition. For the smooth model, however, the peak is shifted slightly to lower temperatures, and the whole curve has a higher value away from the peak. This can be explained by noting that the harmonic springs in the new model introduce additional degrees of freedom not present in the original polymer. Therefore, the specific heat is increased, and the system needs a lower temperature for the freezing transition to occur. In Fig.\,\ref{fig:FEPE}, we have plotted the probability distribution of the energy and the free energy profile at the freezing temperature. The observed barrier height of roughly $19\,k_{\rm{B}} T$ agrees well with the result from Taylor~et\,al. for the pure square-well chain~\cite{tpb_4}.

\subsection{Efficient sampling of reactive pathways}

\begin{figure}
\includegraphics[width=1.0\columnwidth]{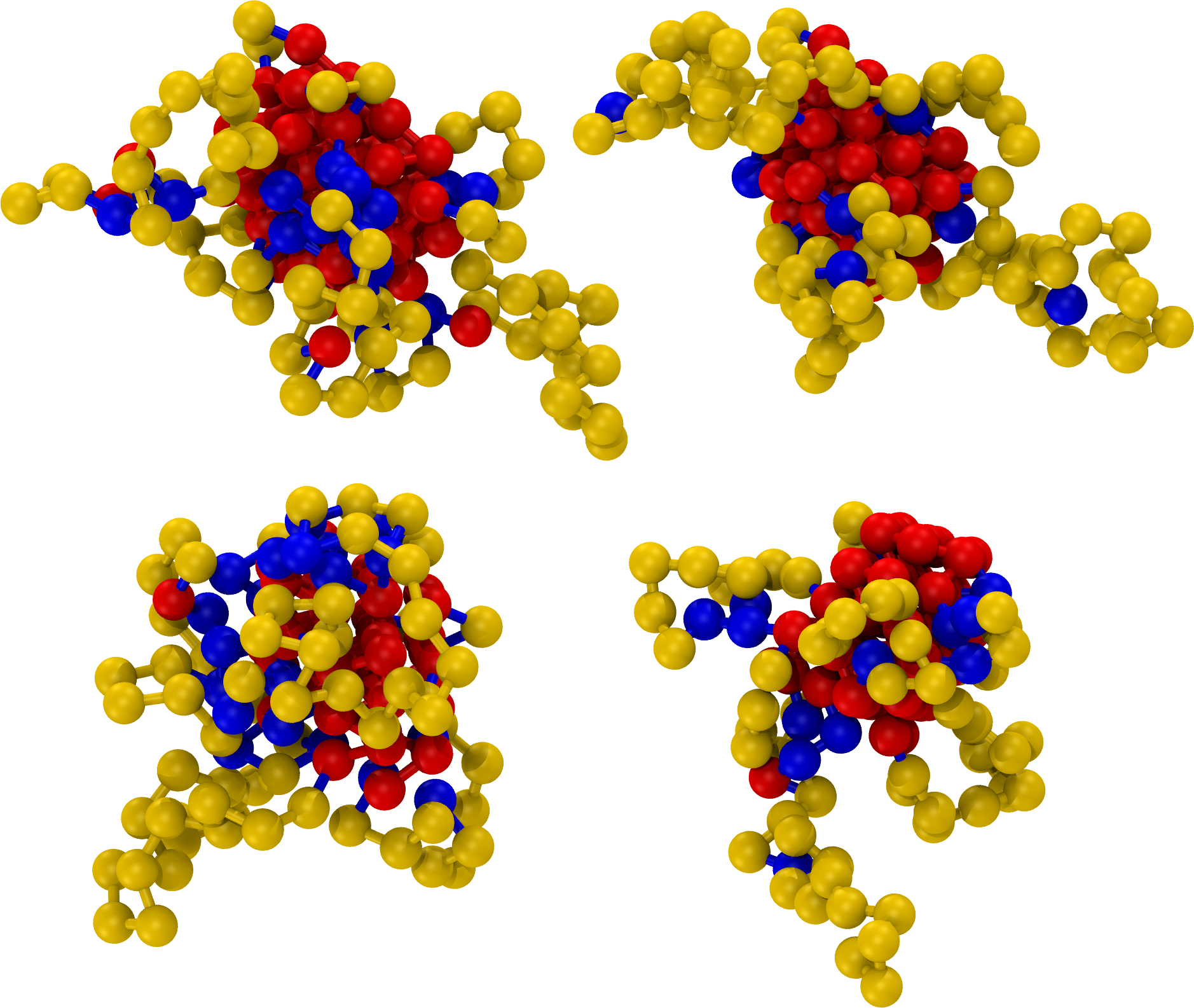}
\caption{(Top) Two shooting point states with a practically identical crystalline core and $c_{XY} = 0.407$. (Bottom) two shooting point states from the same TPS simulation with lower degree of overlap, $c_{XY} = 0.072$. The program \textit{gmmreg}~\cite{gmmreg} has been used for the alignment procedure.}
\label{fig:sp_comparison}
\end{figure}

\begin{figure}
\includegraphics[width=\columnwidth]{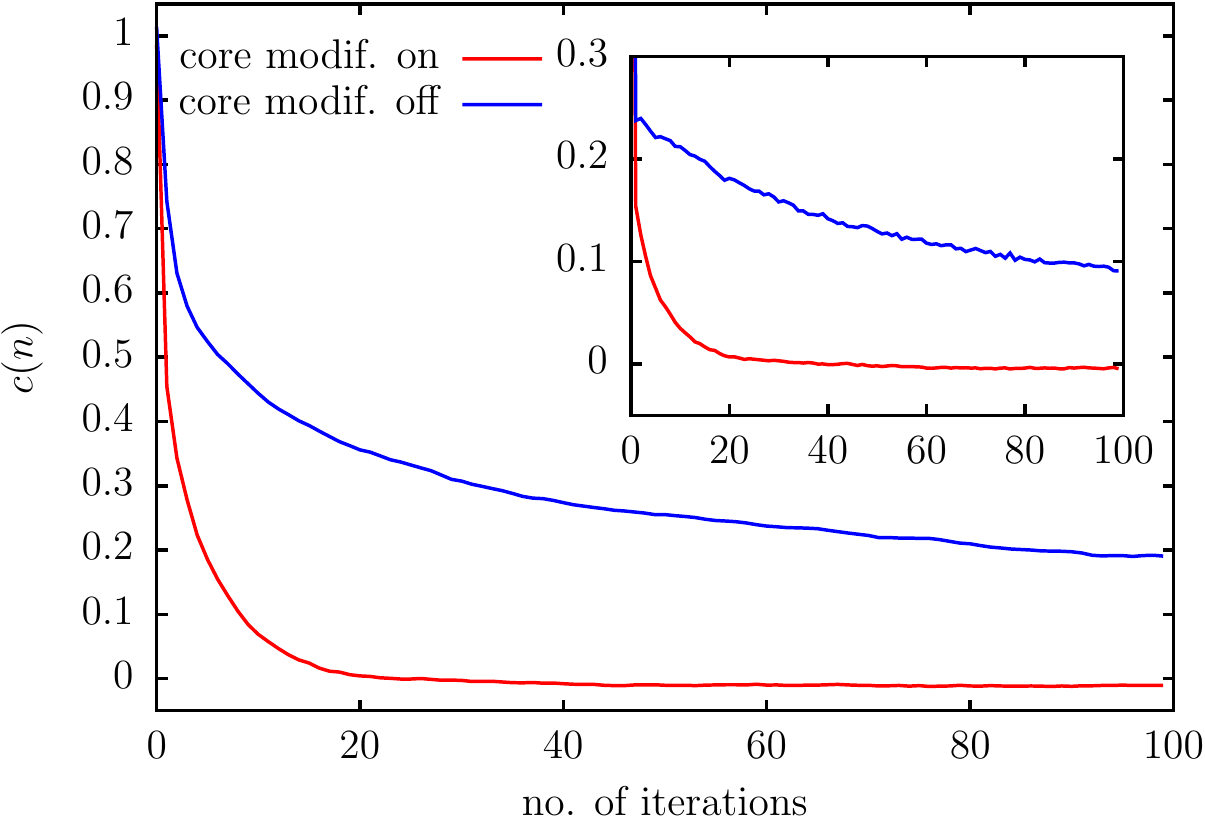}
\caption{Correlation function $c(n)$ as a function of the number of TPS iterations carried out with (red) and without (blue) the use of the core-modification. In the main panel, the correlation function is evaluated at the shooting points and in the inset at the final states of the transition pathways.}
\label{fig:coverlap}
\end{figure}

Some examples of shooting point configurations obtained from a TPS simulation are shown in Fig.\,\ref{fig:sp_comparison}. Even from visual inspection it is clear that the two configurations with a high overlap value share a practically identical crystalline core. For better visibility, the configurations have been spatially aligned using the program \textit{gmmreg}, which implements a point set registration scheme based on gaussian mixture models~\cite{gmmreg}. In Fig.\,\ref{fig:coverlap} we have plotted the correlation function $c(n)$ as a function of TPS cycles as calculated during a TPS run. Correlation functions for the shooting points as well as for the final folded states of the reactive pathways are shown. It can be clearly seen that the use of the core-modification prior to each shooting is vital to de-correlate the state within a reasonable amount of time. Otherwise, even after 100 TPS cycles, there is still a considerable amount of overlap between shooting points and even folded states. In other words, without core modification the simulation is stuck in a single class of similar folding pathways which all lead to final states with identical cores. This is no surprise since already Taylor, Paul, and Binder~\citep{tpb_1} observed that for bigger chains also the Wang--Landau simulations do not converge without the use of bond-bridging moves. Similarly, one must employ bond-bridging moves at the shooting point when performing a TPS simulation in order to overcome the barrier in trajectory space. Note also that in the calculation of $c(n)$, a normalization factor $\langle X^2 \rangle - \langle X \rangle^2$ occurs. It is important to realize that these two averages are different for shooting points and folded states, as there will be more connections present in the folded states compared to the shooting points. Furthermore, when just calculating $\langle X^2 \rangle$ and $\langle X \rangle^2$ from a single time series of shooting points, one will make an error, as these points along a time series are not properly de-correlated. Therefore, we have performed a number of independent TPS runs, each started from already de-correlated paths, to obtain correct values for $\langle X^2 \rangle$ and $\langle X \rangle^2$.

\subsection{Committor analysis and transition state ensemble}

\begin{figure}
\includegraphics[width=\columnwidth]{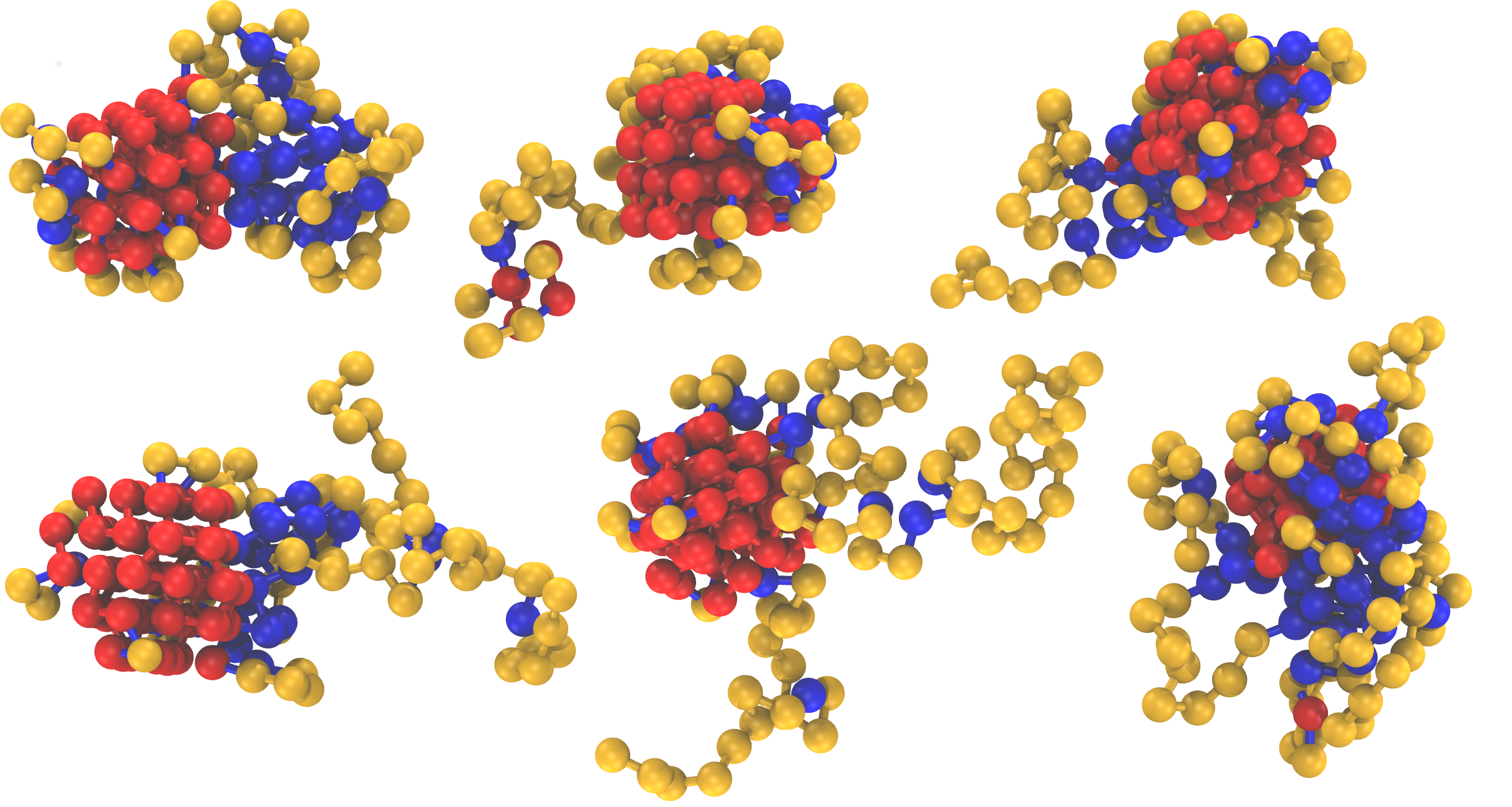}
\caption{Some configurations from the transition state ensemble.}
\label{fig:collage}
\end{figure}

\begin{figure}
\includegraphics[width=\columnwidth]{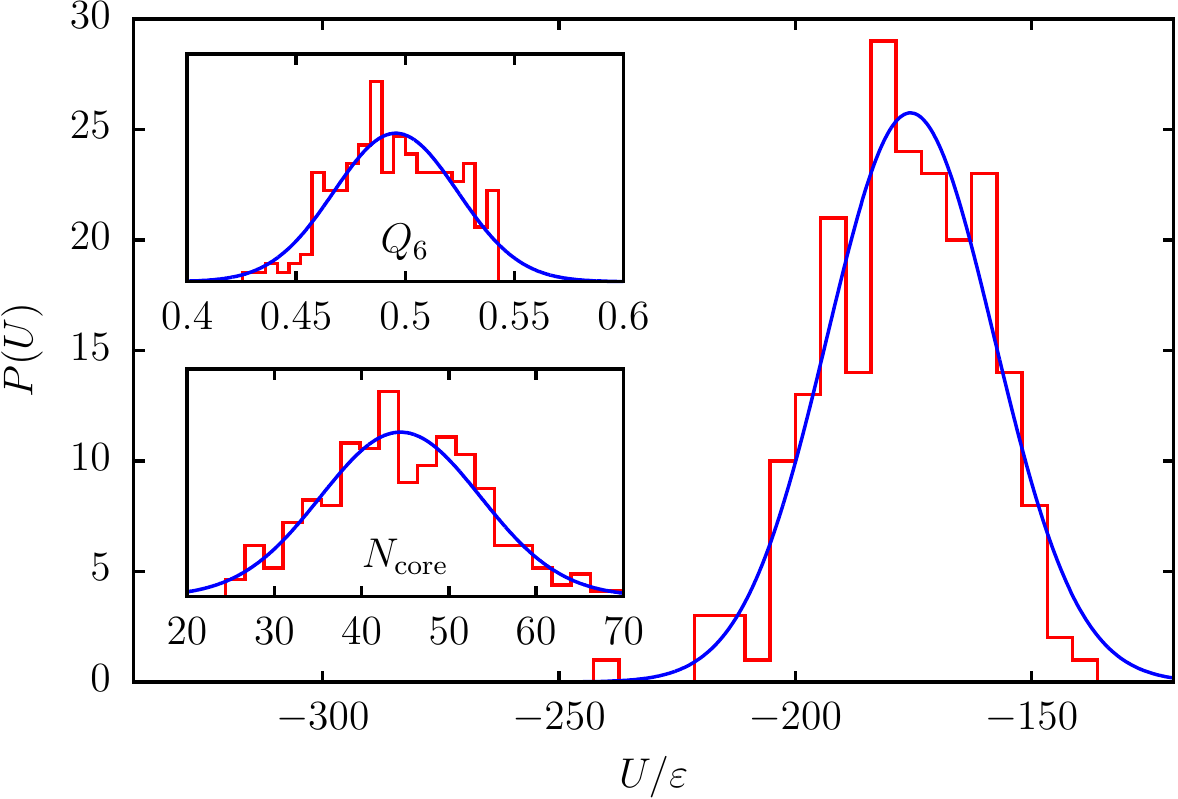}
\caption{Histogram of the potential energy for configurations in the transition state ensemble (red) fitted with a Gaussian distribution (blue). Insets: Same for the global order parameter $Q_6$ (top) and the number of crystalline particles in the core $N_{\rm{core}}$ (bottom).}
\label{fig:ctse_hist}
\end{figure}

\begin{figure}
\includegraphics[width=\columnwidth]{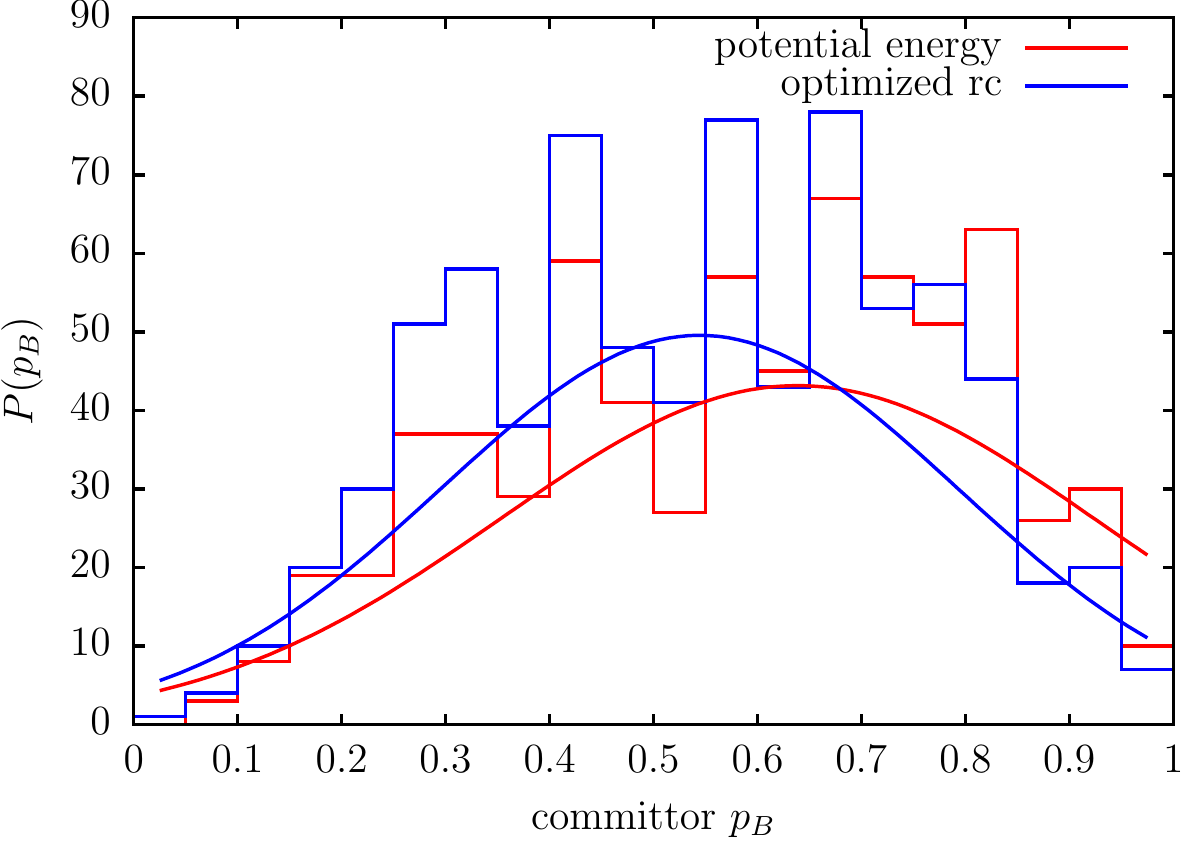}
\caption{(Red) Committor distribution for states with potential energy $U / \varepsilon = -175.6 \pm 5.0$, corresponding to the maximum in Fig.\,\ref{fig:ctse_hist}. (Blue) Committor distribution for states with an value of $r = 0.0 \pm 0.15$ for the optimized reaction coordinate defined in Eq.\,(\ref{eq:rofq}). The total number of states is about the same for both histograms. The solid lines are fits with gaussian functions.}
\label{fig:cpbhist}
\end{figure}

One immediate result of our committor calculations is the transition state ensemble. We define a state to be an element of the TSE if its committor $p_B$ differs from $0.5$ by no more than the statistical uncertainty estimated as
\begin{equation}
\Delta p_B = \sqrt{p_B(1-p_B)/M}.
\end{equation}
Here, $M$ is the number of realized shootings used for the estimation of $p_B$. In total, we have harvested 210~transition states, a few of which are depicted in Fig.\,\ref{fig:collage}. Our first result from analyzing the TSE is an observation made earlier by Taylor, Paul, and Binder~\cite{tpb_2} for the pure square-well chain: typical transition states consist of a crystalline nucleus with one or more chain fragments attached to it. The result  of Taylor et\,al. was drawn from a visual inspection of states which are energetically in between the high-energy coil and the low-energy crystalline phase. Our committor analysis confirms this observation with a more rigorous approach. In Fig.\,\ref{fig:ctse_hist}, we have plotted the distribution of the potential energy, the global order parameter $Q_6$, and the number of crystalline particles in the core for states in the TSE. The core is defined as the largest cluster of crystalline particles. The energies of configurations belonging to the TSE are between $-220\,\varepsilon$ and $-140\,\varepsilon$, roughly corresponding to the barrier region in Fig.\,\ref{fig:FEPE}. The broad distribution seen in Fig.\,\ref{fig:ctse_hist} indicates that the potential energy is not well suited for an accurate description of the progress of the transition. This observation is confirmed by the committor distribution (Fig.\,\ref{fig:cpbhist}) for states from the transition path ensemble with energy $E / \varepsilon = -175.6 \pm 5.0$ corresponding to the peak of the energy distribution shown in Fig.\,\ref{fig:ctse_hist}. Instead of being sharply peaked around a value of $1/2$ as one would expect for a good reaction coordinate, the distribution is broad and includes committor values from $0$ to $1$.

\begin{figure}
\includegraphics[width=\columnwidth]{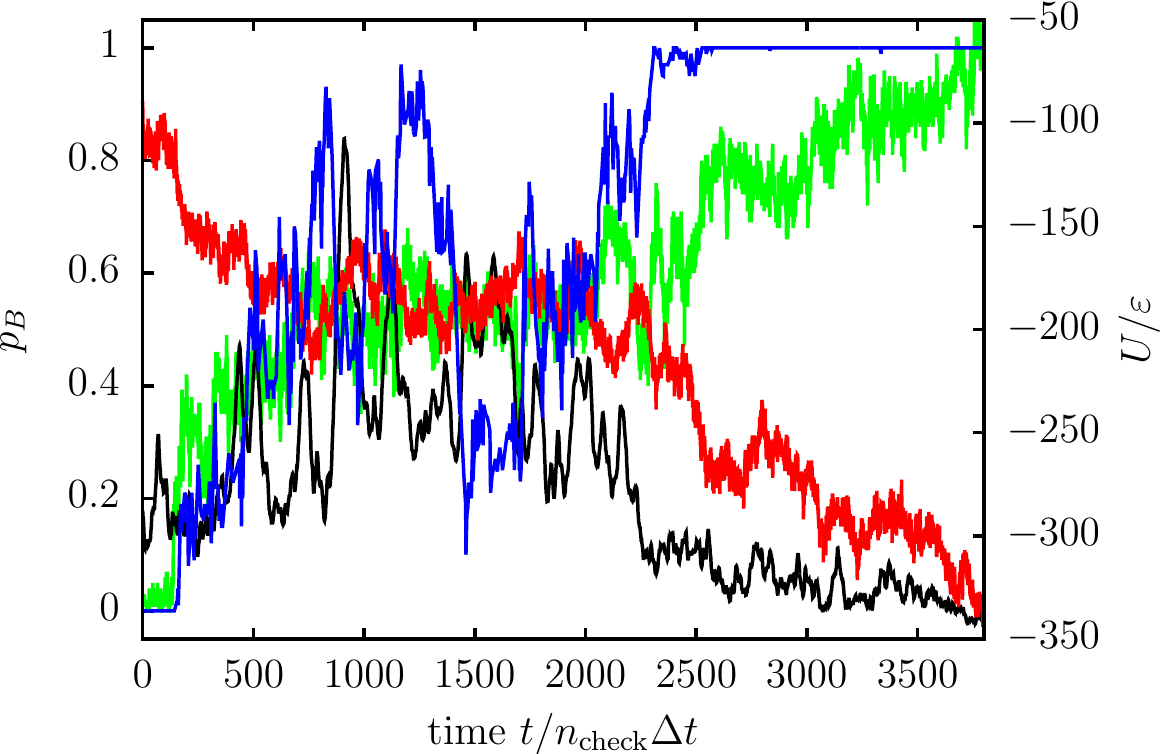}
\caption{Time evolution of the committor $p_B$ (blue), the potential energy $U$ (red), the number of (crystalline) particles in the core $N_{\rm{core}}$ (green, scale not shown), and the radius of gyration $R_g$ (black, scale not shown) along a typical transition path.}
\label{fig:cpboft}
\end{figure}

As seen in Fig.\,\ref{fig:cpboft}, the committor along a typical (folding) transition path is not simply a monotonically increasing function of time. Instead, starting at a value of $0$ at the stable state $A$, it oscillates up and down several times before finally reaching $1$ at the stable state $B$ at the end of the trajectory. This is indicative of the rough and diffusive nature of the freezing transition.

\begin{figure}
\includegraphics[width=\columnwidth]{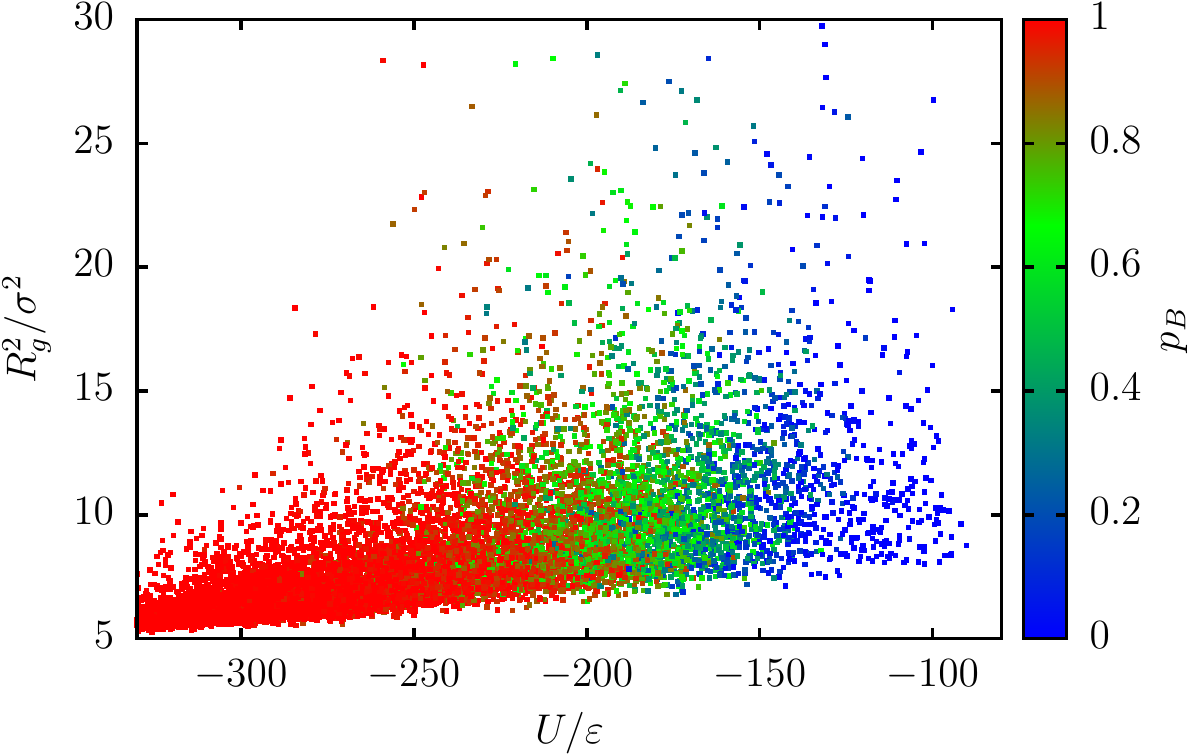}
\caption{Scatter plot of the committor $p_B$ as a function of the potential energy $U$ and the squared radius of gyration $R_g^2$. The dots are colored according to the corresponding committor value according to the color code shown in the bar at the right. Note that in the vicinity of $U / \varepsilon = -200$, there are red dots ($p_B = 1$) next to blue dots ($p_B$ = 0).}
\label{fig:cdotplot_u-rg2}
\end{figure}

From a two-dimensional free energy landscape as a function of the total energy $U$ and the squared radius of gyration $R_g^2$, Taylor~et\,al.~\cite{tpb_4} identified the dominant folding pathway as the minimum free energy path. Our results strongly suggest that such a dominant folding pathway is only representative in an average sense, especially regarding the squared radius of gyration. In Fig.\,\ref{fig:cdotplot_u-rg2} we have plotted the committor values for the configurations from our harvested transition paths in the $U$-$R_g^2$ plane. While the committor is clearly smaller for higher energies and bigger for lower energies, there is almost no structure at all in the $R_g^2$ direction. Also, states with committor values near $0$ and $1$, respectively, can be found next to each other along this axis. This implies that the combination of $U$ and $R_g^2$ is a rather poor choice for predicting committor values and describing reaction pathways. As an illustrative example, one might think of a case where a long coil-like fragment is attached perpendicularly to a crystalline core. The value of $R_g^2$ for such a configuration will be rather large. However, after just one pivot move about one of the particles within the coil-like fragment, $R_g^2$ might tremendously decrease by rotating a significant part of the chain nearer to the core. At the same time, it is rather likely that neither the total energy nor the committor have changed at all after such a move.  

\subsection{Search for a reaction coordinate}

\renewcommand*\arraystretch{1.2}
\begin{table}[htbp]

\begin{tabular}{l|r r r}
\hline
\hline
 & \multicolumn{1}{l}{likelihood 1} & \multicolumn{1}{l}{likelihood 2} & \multicolumn{1}{l}{likelihood 3} \\ \hline
$U$ & \textbf{-501687} & -501687 & -490654 \\
$N_{\rm{core}}$ & -523682 & -495948 & -490654 \\
$Q_6$ & -631293 & \textbf{-490654} & -490654 \\
$\gamma$ & -671208 & -501275 & -490528 \\
$Q_6^{\rm{peri}}$ & -682370 & -500601 & -490654 \\
$R_g$ & -713280 & -498237 & \textbf{-482159} \\
$I_1$ & -735140 & -500141 & -490654 \\
$I_3$ & -735190 & -497896 & -482397 \\
$I_2$ & -752408 & -498003 & -487678 \\
$Q_4^{\rm{peri}}$ & -787806 & -500086 & -490654 \\
$a$ & -798601 & -496548 & -482170 \\
$Q_4$ & -808050 & -495809 & -490649 \\
$Q_4^{\rm{core}}$ & -818773 & -499298 & -490580 \\ 
$Q_6^{\rm{core}}$ & -826644 & -496584 & -490554 \\ \hline 
BIC & 1003386 & 981326 & 964342 \\ \hline \hline
\end{tabular}
\caption{likelihood 1: log likelihood scores for the different single collective variables. likelihood 2: log likelihood scores for the combination of $U$ with the given variable, where the optimization was performed over the coefficients for both $U$ and the other variable. likelihood 3: the same for the combination of ($U$, $Q_6$) with the given variable. The maximum for each combination is highlighted in bold type. Also the Bayesian information criterion for each combination with maximum likelihood is given. Note that smaller BIC values are better.}
\label{tab:likelihoodscores}
\end{table}

The main objective of the search for a reaction coordinate is to find a -- preferably simple and transparent -- function of the system variables that encodes the progress of the transition and reduces many correlated degrees of freedom to a single, essential one. However, it is clear from Figs.\,\ref{fig:ctse_hist}~and~\ref{fig:cpboft}, that neither the total potential energy $U$ nor the number of particles in the crystalline core $N_{\mathrm{core}}$ alone is able to serve as a reaction coordinate. Nonetheless, one might hope that a combination of these two quantities, possibly including some additional physical parameters, could lead to better results.

\begin{figure}
\includegraphics[width=\columnwidth]{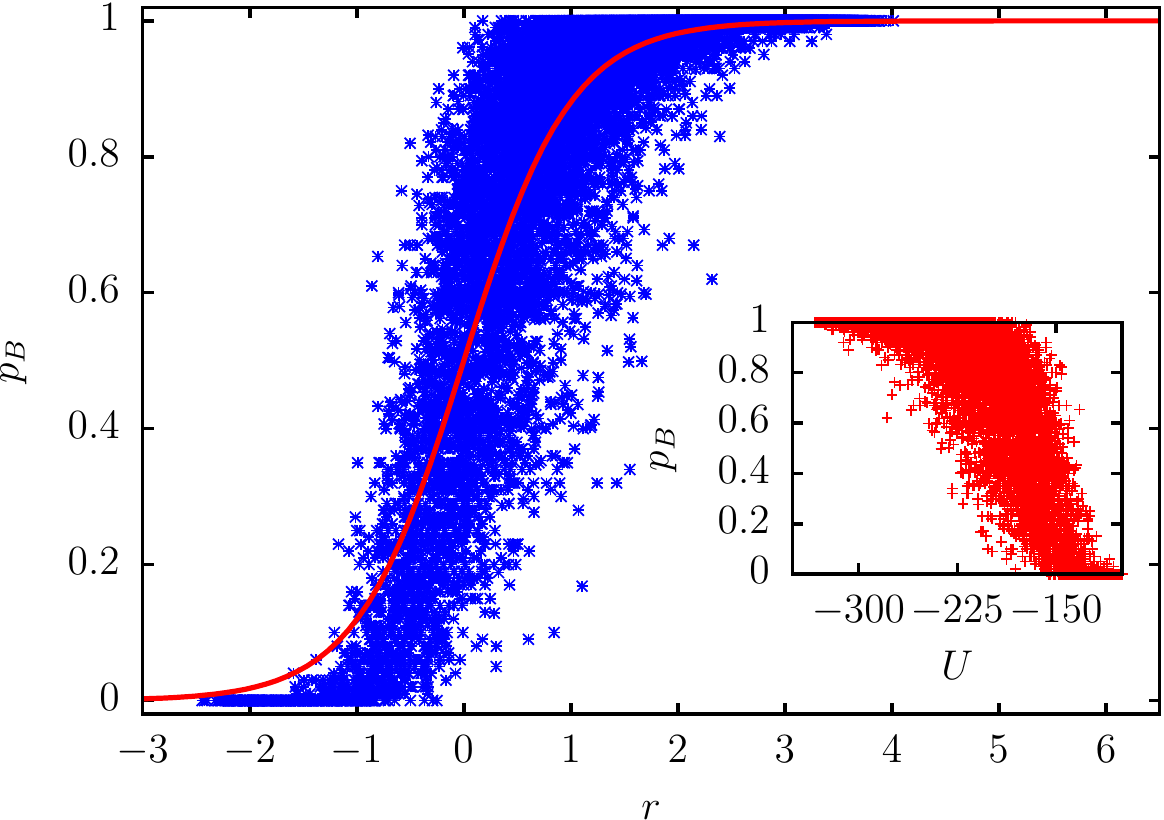}
\caption{Committor $p_B$ as a function of the optimum reaction coordinate $r$, which is a linear combination of the potential energy $U$ and the global order parameter $Q_6$. The red line is the model reaction coordinate of Eq.\,\eqref{eq:pbofr}. Inset: Committor as a function of $U$ alone.}
\label{fig:cpb_likelihood_exptanh}
\end{figure}

We have tested the quality of several collective variables as a reaction coordinate by computing the likelihood $L(\alpha)$ from Eq.\,(\ref{eq:Lalpha}) for optimized coefficients $\alpha$. It turns out that when one uses only one variable as a model reaction coordinate, $U$ gives the best likelihood (see Tab.\,\ref{tab:likelihoodscores}). Furthermore, when including an additional variable and performing the optimization over both $U$ and the second variable, the use of the global order parameter $Q_6$ gives the best improvement. The improvement is better than with the inclusion of $N_{\mathrm{core}}$, even though, for single variables, $N_{\mathrm{core}}$ has a higher likelihood score. However, since the energy is dominated by the number of non-permanent bonds, and most bonds occur in the crystalline core, it is clear that $U$ and $N_{\mathrm{core}}$ are highly correlated. Therefore, $N_{\mathrm{core}}$ can add only very little information that is not already present in $U$, and consequently, the variable pair ($U$, $N_{\rm{core}}$) gets a lower likelihood score than ($U$, $Q_6$). Iterating the procedure one step further, we find that including the radius of gyration $R_g$ improves the likelihood the most. However, the relative improvement for a third variable is rather small, confirming our previous observation that the radius of gyration does not carry much information about the crystallization mechanism. Furthermore, both the anisotropy $a = I_3 / I_1 - 1$ of the configuration, with the moments of inertia $I_3 = I_{\rm{max}}$ and $I_1 = I_{\rm{min}}$, as well as $I_3$ alone, give a very similar likelihood score when used as a third variable in the optimization. In order to test the robustness of the procedure, we have performed the optimization procedure for smaller subsets of the full data set of known committor values. While $U$ and $Q_6$ are ranked first and second consistently, the third variable is often different for each run. For example, in one case $I_2$, $I_3$, and $a$ were equally well suited as a third variable, giving almost identical likelihood scores. This implies that a third variable in addition to the combination ($U$,$Q_6$) does not improve the quality of the optimized reaction coordinate in a significant way. The log likelihood scores for the full dataset are listed in Table\,\ref{tab:likelihoodscores}. Other collective variables considered in the optimization were the order parameters $Q_4$ and $Q_6$ evaluated for core ($Q_4^{\text{core}}$,$Q_6^{\text{core}}$) and peripheral particles ($Q_4^{\text{peri}}$,$Q_6^{\text{peri}}$) only, and the biggest eigenvalue $\gamma$ of the contact matrix (as used by Růžička and coworkers~\citep{allen_folding}). The contact or Laplacian matrix is defined as
\begin{equation}
\mathbb{G}_{ij} = 
\begin{cases}
-1  & |i-j| > 1 \quad \mathrm{and} \quad R_{ij} \leq \lambda \sigma, \\
0 & |i-j| > 1 \quad \mathrm{and} \quad R_{ij} > \lambda \sigma, \\
0 & |i-j| = 1, \\
- \sum_{k \neq j} {G}_{kj} & |i-j| = 0.
\end{cases}
\label{eq:contactmatrix}
\end{equation}
In our work, we have actually used the matrix of pair energies instead, which is very similar to the above definition due to the shape of the pair potential. Note that as observed by Růžička and coworkers previously, it makes little difference for the behavior of $\gamma$ whether one uses the above definition for the diagonal elements of the matrix, or sets them to zero as it is the case when using the pair energies. The committor as a function of the optimal reaction coordinate constructed from $U$, $Q_6$ and $R_g$ is shown in Fig.\,\ref{fig:cpb_likelihood_exptanh}. Although this reaction coordinate yields the best likelihood score, there remains a considerable spread in the committor for any particular value of the reaction coordinate.

In Fig.\,\ref{fig:cpbhist}, we have plotted the committor distribution for all states with a value of $r = 0.0 \pm 0.15$ for the optimized reaction coordinate. For both the potential energy and the optimized RC, the distribution has a small trough around $p_B = 1/2$. Since all states were selected at random, a likely explanation is that the system simply does not spend a long time at committor values around $1/2$. If the proposed reaction coordinate is of poor quality, the committor distribution for the transition value of the RC will look similar to the unrestricted distribution, which also has a trough around $1/2$. However, compared to the distribution for potential energy values of $U / \varepsilon = -175.6 \pm 5.0$, for the optimized RC one sees a slight improvement because the width of the histogram has decreased due to the optimization, while the peak of the distribution is shifted towards a value of $0.5$. In particular, Gaussians fitted to the histograms have a mean of $0.64$ with a standard deviation of $0.29$ for the potential energy as reaction coordinate and a mean of $0.54$ and a standard deviation of $0.25$ for the optimized reaction coordinate.

\section{Discussion}
\label{sec:discussion}

For sufficiently small interaction ranges $\lambda$, the square-well polymer chain shows a two-state folding transition from the extended coil directly to the crystalline state. Using TPS simulations with a new core modification move, which has proved crucial for the ergodic sampling of transition pathways, we have studied the folding mechanism in detail. A committor analysis of the harvested transition pathways has confirmed the earlier observation that transition states have a single crystalline nucleus, while the rest of the system is still in a coil-like configuration. While the fully crystalline state and the coil state can be distinguished based on the total potential energy, which essentially counts the number of close contacts between monomers, the energy is not well suited as reaction coordinate. The potential energy of transition states is distributed around $U / \varepsilon = -176$, which also coincides with the maximum of the free energy curve at the transition temperature. However, the broad distribution of committor values at $U / \varepsilon = -176$ shows that the potential energy does not accurately quantify the progress of the crystallization transition. Using the likelihood maximization method of Peters and Trout~\cite{peters_rc}, we have constructed an optimized reaction coordinate, which is a linear combination of the the potential energy and the global order parameter $Q_6$ of the polymer. This optimized reaction coordinate captures the progress of the folding transition more accurately than the energy alone or any other combination of two variables we tested, indicating that variables describing the structure and the overall shape of the polymer are needed in order to understand the transition mechanism. However, the improvement of the reaction coordinate obtained by including $Q_6$ is marginal.

The question arises whether the optimum reaction coordinate identified for the specific model studied in this paper is transferable to other polymer models. In our model with short-range attractions and strongly repulsive cores, the potential energy works as order parameter because it is proportional to the number of contacts. In the case of polymers with additional contributions to the potential energy, such as torsional and angular potentials, it will be more appropriate to use the number of contacts directly rather than the potential energy. Similarly, the order parameter $Q_6$ is sensitive to a close-packed structure, which occurs in the crystalline state of the square-well chain. For more complex polymers, it is likely that one will get better results when using an order parameter which is sensitive to the particular ground state structure of the system under consideration.

An interesting aspect of of the transition has been noted recently by Růžička, Quigley and Allen~\cite{allen_folding} in forward flux sampling simulations of a slightly modified version of the chain allowing for the application of collision dynamics. They have analyzed the distribution of the largest eigenvalue $\gamma$ of the polymer's Laplacian matrix. This variable exhibits a different distribution depending on the folding probability. More specifically, folding pathways with a high folding probability, as well as unfolding pathways, show a bimodal distribution of $\gamma$, while the equilibrium distribution for the same temperature is unimodal. The second peak occurs at a higher value of $\gamma$, which also corresponds to configurations with an already ordered crystalline nucleus.

While $\gamma$ carries a signature characteristic for the two stable states, it does not convey much useful information about the progress of the transition. The features in the distribution of this variable~\citep{allen_folding} have a straightforward explanation: Configurations which have a finite folding probability already need to have some degree of crystallinity, otherwise they will not fold even if their energy is rather low. In other words, while $\gamma$ carries enough information to decide whether there is \textit{some} or \textit{zero} folding probability, the variable cannot be used to make an accurate prediction of the actual committor value if it is anything other than zero, and, therefore, does not perform well as a reaction coordinate. We have confirmed this result even when we include several or even all eigenvalues of the contact matrix in the reaction coordinate approximation. A possible explanation is that even the full set of eigenvalues suffers from the same flaw as $\gamma$ alone: Similar to other global order parameters, they are only a measure of the overall crystallinity present in the system. By construction, they are completely symmetric under reordering of the particles. However, in a polymer chain, which has linked neighbors, the order of the monomers actually does matter, a fact that is completely neglected when using such measures of crystallinity. Therefore, it remains a challenging task to construct better order parameters for homopolymers, which take into account the actual order along the chain, while still being symmetric under other operations, such as reversing the labeling without changing the connectivity. More elaborate machine learning approaches such as support vector machines may be helpful in this endeavor.

\begin{acknowledgments}
We thank Clemens Moritz for suggesting the use of the bond-bridging move in TPS simulations. The Initiativkolleg “Computational Science” of the University of Vienna and the Austrian Science Fund (FWF) within the SFB ViCoM (grant no. F41) is gratefully acknowledged for financial support. The computational results presented have been achieved using the Vienna Scientific Cluster (VSC).
\end{acknowledgments}

\appendix

\section{Bond-bridging move with variable bond lengths}

\label{ap:bond_bridging}

\begin{figure}
\includegraphics[width=0.95\columnwidth]{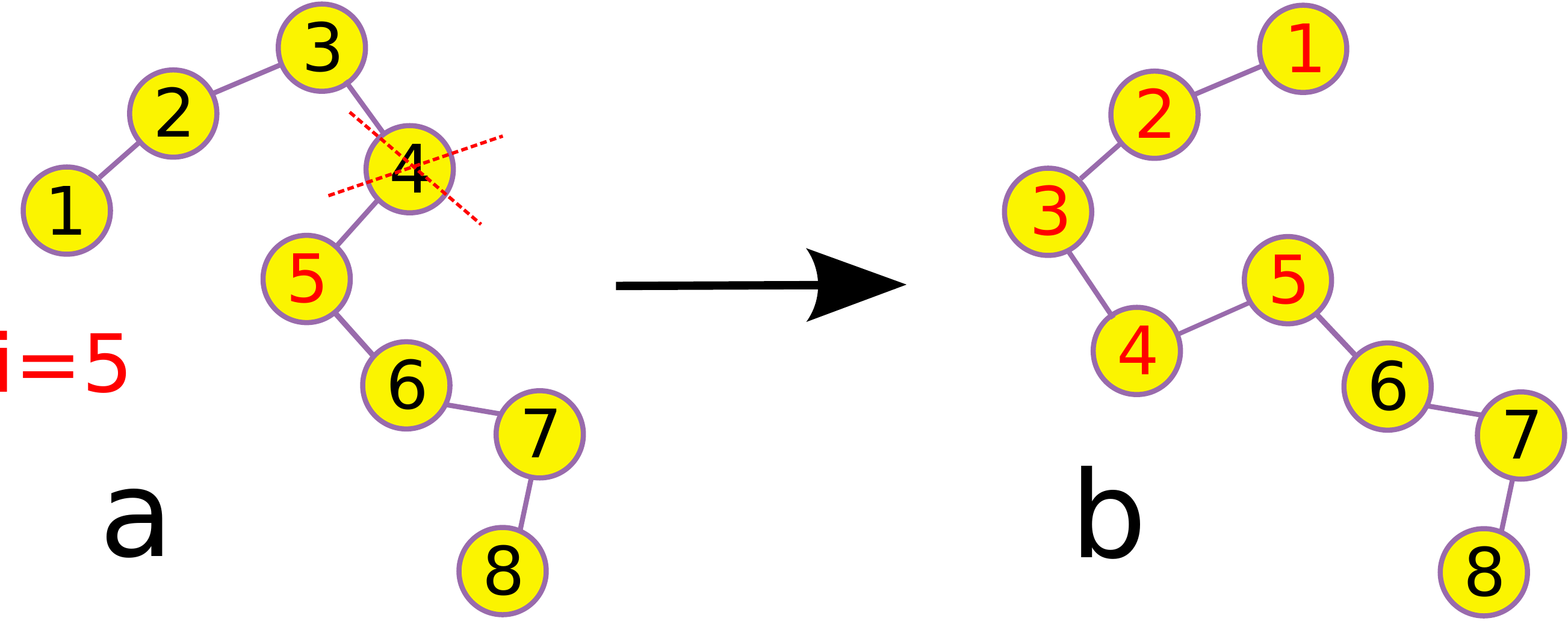}
\caption{Schematic representation of the bond-bridging move.}
\label{fig:state_a}
\end{figure}

The bond-bridging move used in our TPS simulation works as follows. First, end particle $1$ or $N$ is chosen with equal probability. Then, all interior monomers ($i > 3$ or $i < N - 2$) within a distance of $2 \, L$ of the chosen end are identified, and one of these, denoted by $i$, is chosen at random. This monomer is reconnected to the end via removal and re-insertion of the next chain neighbor in the direction of the end, namely particle $i-1$ or $i+1$. A schematic view of this procedure is given in Fig.\,\ref{fig:state_a}. The acceptance probability for the move is
\begin{equation}
P_{\text{acc}}(a \rightarrow b) = \min \left[1, \frac{\rho(b)}{\rho(a)} \frac{b_a}{b_b} \frac{R_b}{R_a} \right].
\end{equation}
Here, $b_a$ ($b_b$) stands for the number of possible bridging partners present in state $a$ ($b$) and $R_a$ is the distance of monomer $i$ to the previously selected chain end, with $R_b$ defined accordingly. $\rho$ is the equilibrium distribution to be sampled. As usual, the value of $\rho(b) / \rho(a)$ depends on the type of simulation that is performed: We have $\rho(b) / \rho(a) = e^{-\beta(E_b - E_a)}$ for a canonical ensemble at inverse temperature $\beta$ and $\rho(b) / \rho(a) = g(E_a) / g(E_b)$ in the case of a Wang--Landau simulation.

As we will discuss in the following, particular care has to be taken in deriving the acceptance probability, because in our modified square-well polymer, bond lengths are allowed to fluctuate. The move is designed such that none of these distances are changed, in other words, the change in the potential energy from the harmonic springs is always zero. 

\begin{figure}
\includegraphics[width=0.45\columnwidth]{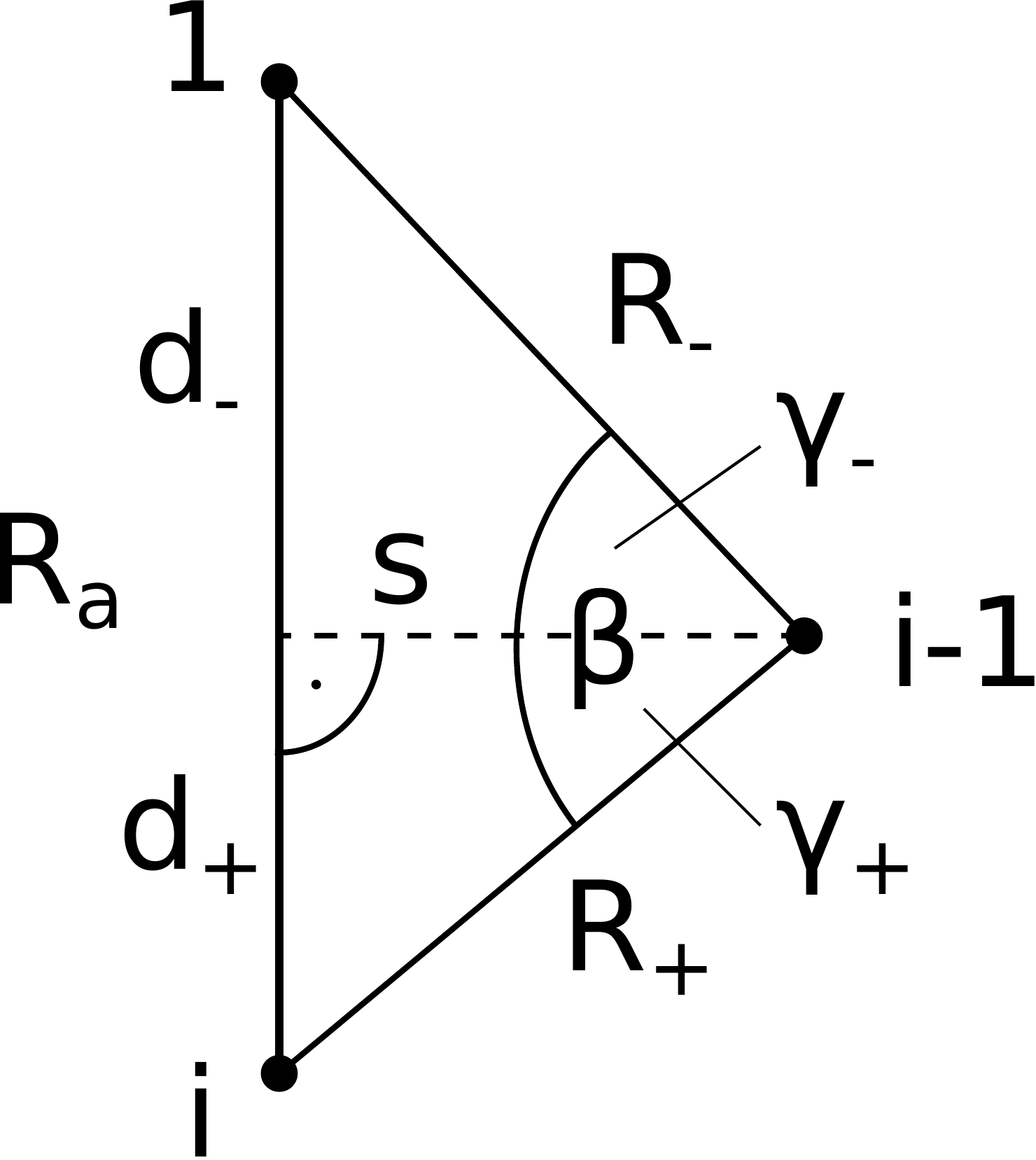}
\caption{Distances and angles involved in the bond-bridging move. Here, particle $1$ has been chosen as the end to be reconnected, therefore particle $i-1$ is re-inserted between particle $1$ and particle $i$.}
\label{fig:bondbridging_distances}
\end{figure}

\begin{figure}
\includegraphics[width=0.35\columnwidth]{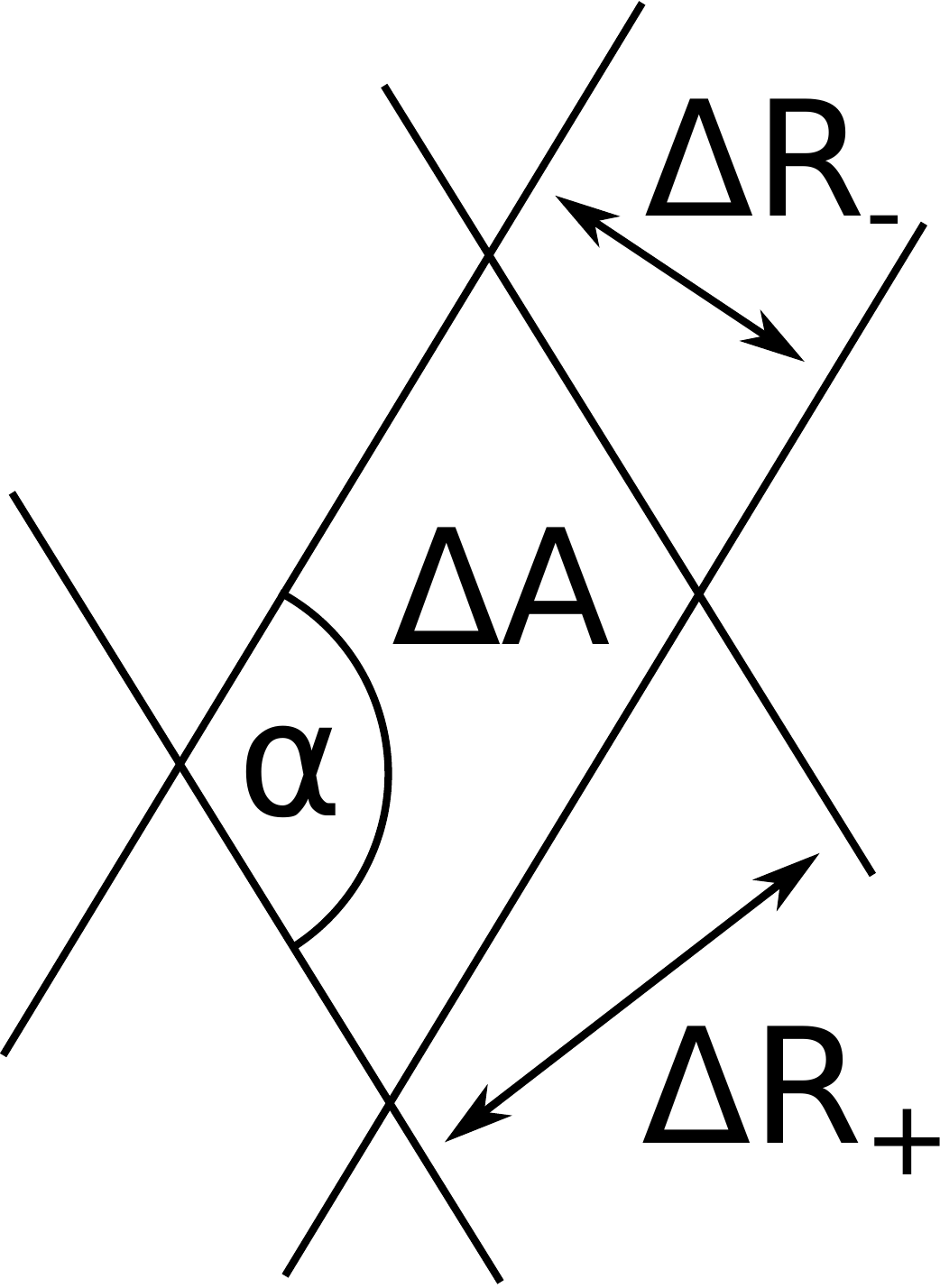}
\caption{Close-up of the geometry at the particle insertion point.}
\label{fig:bondbridging_parallelogram}
\end{figure}

In all the following, we will assume that particle $1$ has been selected as the end to be reconnected. The situation for end particle $N$ is of course identical, but $i-1$ has to be replaced by $i+1$ and $i-2$ by $i+2$~etc. Let us also denote the distances to keep fixed with $R_+ = |\vec{R}_i - \vec{R}_{i-1}|$ and $R_- = |\vec{R}_{i-1} - \vec{R}_{i-2}|$. The chain is then re-connected as indicated in Fig.\,\ref{fig:bondbridging_distances}. We have $R_a = d_+ + d_-$, and can calculate
\begin{equation}
d_- = \frac{R_a^2 - R_+^2 + R_-^2}{2 R_a}.
\end{equation}
Note that due to the variable bond lengths, it is possible that $R_+ + R_- < R_a$. In that case, there is no possibility to perform the re-connection, and the move is rejected. Detailed balance for the move can be satisfied with the usual Metropolis acceptance criterion
\begin{equation}
P_{\text{acc}}(a \rightarrow b) = \min \left[ 1, \frac{\rho(b)}{\rho(a)} \frac{P_{\text{gen}}(b \rightarrow a)}{P_{\text{gen}}(a \rightarrow b)} \right],
\end{equation}
where $\rho(x)$ is the equilibrium distribution to be sampled and $P_{\text{gen}}(a \rightarrow b)$ is the probability to generate configuration $b$ out of $a$. Hence, we need to know the generation probability in order to get the correct acceptance rule for the bond-bridging move. The generation probability can be written as a product of three factors:
\begin{equation}
P_{\text{gen}}(a \rightarrow b) = \frac{1}{2} \frac{1}{b_a} \frac{1}{n(b)}.
\end{equation}
Here, $1/2$ arises from the fact that there are two ends to choose from and $b_a$ is the number of possible bridging partners present in state $a$. Furthermore, $n(b)$ is the number of possible configurations for $b$ once the other choices have already been made. This number is proportional to the configuration space volume available for the choice of $b$ under the imposed constraints: $n(b) = c \Delta v(b)$. In order to estimate $\Delta v(b)$ for the given geometry, we have to realize that a constraint on a distance is constructed using a delta function in the distribution function. In other words, $\delta(R - R')$ actually means that $R' < R < R' + \Delta R$, where $\Delta R$ is infinitesimal. In the case of our move, the distances to be kept fixed are $R_+$ and $R_-$. In Fig.\,\ref{fig:bondbridging_parallelogram}, we have illustrated the geometry at the re-insertion point, with the parallel lines indicating the infinitesimal constraints. The available configuration space volume is proportional to the (also infinitesimal) cross section $\Delta A$. Since the azimuthal angle of the re-insertion is random, this has to be multiplied with the circumference of the circle defined by the rotation of the re-insertion point around the axis from $1$ to $i$:
\begin{equation}
\Delta v = 2 \pi s \Delta A.
\end{equation}

The actual value for $\Delta A$ can be calculated by looking at the geometry at the re-insertion point (Fig.\,\ref{fig:bondbridging_parallelogram}). We have
\begin{equation}
\Delta A = \frac{\Delta R_+ \Delta R_-}{\sin \alpha}.
\end{equation}
Also, $\alpha = \pi - \beta$, therefore $\sin \alpha = \sin \beta$. To get an expression that only includes known distances, we first split the triangle defined by the positions of particles $i$, $i-1$ and $1$ into two right triangles: $\beta = \gamma_+ + \gamma_-$. Hence,
\begin{align}
\sin \beta &= \sin( \gamma_+ + \gamma_- ) \nonumber \\
 &= \sin \gamma_+ \cos \gamma_- + \sin \gamma_- \cos \gamma_+.
\end{align}
The sines and cosines can now be expressed as ratios of known lengths:
\begin{align}
\sin \gamma_+ &= \frac{d_+}{R_+}, &
\sin \gamma_- &= \frac{d_-}{R_-}, \nonumber \\
\cos \gamma_+ &= \frac{s}{R_+}, &
\cos \gamma_- &= \frac{s}{R_-}.
\end{align}
Putting all together, we get
\begin{equation}
\Delta v = \frac{2 \pi \Delta R_+ \Delta R_- R_+ R_-}{R_a}.
\end{equation}
Hence the generation probability is
\begin{equation}
P_{\text{gen}}(a \rightarrow b) = \frac{1}{b_a} \frac{R_a}{4 \pi \Delta R_+ \Delta R_- R_+ R_-}.
\end{equation}
This implies
\begin{align}
& P_{\text{acc}}(a \rightarrow b) = \min \left[ 1, \frac{\rho(b)}{\rho(a)} \frac{b_a}{b_b} \times \right. \nonumber \\
& \qquad \left. \frac{c 4 \pi \Delta R_+ \Delta R_- R_+ R_-}{R_a} \frac{R_b}{c 4 \pi \Delta R_+ \Delta R_- R_+ R_-} \right] \nonumber \\
& \hphantom{P_{\text{acc}}(a \rightarrow b)} = \min \left[1, \frac{\rho(b)}{\rho(a)} \frac{b_a}{b_b} \frac{R_b}{R_a} \right].
\end{align}
Note that this result holds regardless of whether $R_+ = R_-$ or not. In other words, we can use the same acceptance criterion as for the case of fixed bond lengths.


\end{document}